# Hybrid Electrothermal Simulation of a Three-Dimensional Fin-Shaped Field-Effect Transistor Based on GaN Nanowires

Qing Hao*, Hongbo Zhao, Yue Xiao, Michael Brandon Kronenfeld

Department of Aerospace and Mechanical Engineering, University of Arizona

Tucson, AZ 85721-0119

Electronic mail: qinghao@email.arizona.edu

Abstract

In recent years, three-dimensional GaN-based transistors have been intensively studied for their dramatically improved output power, better gate controllability, and shorter channels for speedup and miniaturization. However, thermal analysis of such devices is often oversimplified using the conventional Fourier's law and bulk material properties in thermal simulations. In this aspect, accurate temperature predictions can be achieved by coupled phonon and electron Monte Carlo simulations that track the movement and scattering of individual phonons and electrons. However, the heavy computational load often restricts such simulations to nanoscale devices, while a real chip is of millimeter to centimeter sizes. This issue can be addressed by a hybrid simulation technique that employs the Fourier's law for regions away from the hot spot. Using this technique, accurate electrothermal simulations are carried out on a nanowire-based GaN transistor to reveal the temperature rise in such devices.

With their high radio-frequency power density, operation frequency, and breakdown voltage, GaN-based devices exhibit significant advantages over silicon-based devices for high-power and high-frequency applications. Conventionally, tremendous efforts have been dedicated to GaN devices based on the two-dimensional electron gas (2DEG) on the heterojunction between planar films of GaN and $Al_xGa_{1-x}N$ alloy.[1] Beyond such layered device structures, GaN fin-shaped field effect transistors (FinFETs) have also been studied in recent years, using GaN nanowires[2] or heterojunctions between GaN and its alloys.[2-5] With better gate controls, FinFETs reduce the detrimental short-channel effects and thus allow further miniaturization for high-speed applications. In a recent study, the maximum drain current can reach 1.1 A/mm for GaN/AlGaN FinFETs, compared with 0.37 A/mm current in a reference planar GaN/AlGaN device.[5]

As a general concern for the performance of high-power GaN devices, overheating of such devices can dramatically deteriorate the device performance and shorten the lifetime. In this aspect, electron transport analysis must be coupled with thermal simulations to address the impact of device self-heating. Despite numerous electrothermal studies on conventional planar GaN devices, electrothermal simulations are rare for GaN-based FinFETs. More importantly, conventional Fourier's law and bulk material thermal conductivities are often used in existing electrothermal simulations, as reviewed in a recent work.[6] This treatment is invalid at micro- to nano-scales, where the phonon mean free paths (MFPs) become comparable or longer than the structure sizes. In this case, phonon Boltzmann transport equation (BTE) should be solved together with electron simulations. Such calculations often involve a heavy computational load, particularly when heat spreading across the sub-millimeter chip is further considered.

In practice, detailed electron and phonon transport within a GaN transistor and heat spreading across the whole chip can be both incorporated using a hybrid simulation technique.[6,7]

In such simulations, coupled electron and phonon Monte Carlo (MC) simulations are used to predict the temperature rise for the transistor regions. These MC simulations track the movement and scattering of individual electrons and phonons and can statistically obtain the solution for the phonon and electron BTE. Complicated 3D structures and energy dependence of carrier scattering and transport can all be considered in such simulations. For regions away from the hot spot, phonons are anticipated to be in thermal equilibrium with the local temperature so that the Fourier's law analysis becomes valid. In this case, the phonon MC simulation for the transistor region is coupled with the Fourier's law analysis away from the transistor to provide the temperature distribution across the whole chip. This hybrid technique allows accurate temperature predictions of general nanoelectronic devices and is employed here for 3D GaN FinFETs.

For Si-doped GaN nanowires with 120 nm cross-section dimension, bulk phonon dispersion and electronic band structures are still assumed. For electrons, the lowest three conduction band valleys of wurtzite GaN are considered in electron MC simulations, known as the $\Gamma_1$, U, and $\Gamma_3$ valleys.[8] The electronic band structure is described by the analytical Kane's model, $E_i(1+\alpha_i E_i) = \hbar^2 k^2/2m_i$, where $\alpha$ is band nonparabolicity, $E$ is the kinetic energy of electrons, and $i$ is index for the three valleys. The effective mass $m_i$ is evaluated at the bottom of each valley and the energy-dependent effective mass is given as $m_i(1+2\alpha E)$.[9] The electron scattering mechanisms include ionized impurity scattering, polar optical phonon scattering, acoustic deformation potential scattering, and intervalley optical phonon scattering. It is assumed that all Si dopants are activated, which is the case for a high growth or annealing temperature.[10] The expressions for different scattering rates can be found elsewhere.[7] All employed parameters are provided in Table I. In electron MC simulations, the computational domain is half of the nanowire (cut by dot-dash line A in Fig. 1) and is divided into 3×6×100 subcells to count the local electron concentration. The three-dimensional electric field is thus updated during the simulation by solving the Poisson equation with applied terminal voltages and counted electron concentrations inside each subcell.[7] In steady states, both the current and electric field are no longer changed and the energy exchange between hot electrons and phonons are counted within each subcell for the following thermal simulations.

For phonon transport, three identical acoustic phonon branches are considered. In the GaN nanochannel, hot electrons first pass their energy to the topmost longitudinal optical (LO) phonon branch that is fixed at 91.2 meV. These non-propagating hot LO phonons then transfer the energy to acoustic phonons to spread out the heat across the whole device. Assuming three identical sine-shaped acoustic branches, the phonon angular frequency ω is related to the wave vector $q$ by $\omega = \omega_{\max}\sin(\pi q/2q_0)$, in which $\omega_{\max}$ and $q_0$ are the maximum $\omega$ and $q$ value, respectively. Here $q_0$ and the equivalent atomic distance $a_D$ can be computed by $q_0 = \frac{\pi}{a_D} = (6\pi^2 N)^{1/3}$, with $N$ as the volumetric density of primitive cells. The maximum angular frequency can be calculated from $a_D$ as $\omega_{\max} = \frac{2v_s}{a_D}$. The essential phonon scattering mechanisms include impurity scattering and the Umklapp process of the phonon-phonon scattering. The overall phonon relaxation time $\tau(\omega)$ is given as $1/\tau(\omega) = A\omega^4 + B_1\omega^2 T\exp(-B_2/T)$, where the first term on the right side is for impurity scattering and the second term is for Umklapp scattering. Parameters used for all materials are obtained by fitting measured bulk thermal conductivities and are listed in Table II. In comparison,[6] the obtained phonon MFP distributions for undoped bulk materials are consistent with existing measurements on bulk SiC and GaN.[11] For Si-doped GaN nanowires, the impurity scattering of phonons is stronger than that in pure GaN and $A$ should be increased. Here $A$ is estimated as $A = \frac{\Gamma V_0}{4\pi v_s^3}$,[12] where the averaged sound velocity among all acoustic branches is $v_s$

=3338 m/s, and the unit volume $V_0$ for wurtzite GaN is 11.42 Å$^3$.[13] For simplicity, only the mass difference due to substitution Si atoms is considered[14] and $\Gamma = f_{Si}\left\{\left(1-\frac{M_{Si}}{\bar{M}}\right)^2\right\}$,[15] where $f$≈4.5×10$^{-5}$ is the fractional concentration of Si atoms at the 4.0×10$^{18}$ cm$^{-3}$ doping level, $M_{Si}$ is the atomic mass of Si, $\bar{M}$ is the averaged atomic mass. The obtained $A$=1.2×10$^{-46}$ s$^3$ is negligible compared with A=5.26×10$^{-45}$ s$^3$in Table II, the latter of which was fitted for real GaN samples with unintentional defects.

For GaN doped with Si, phonon scattering by free electrons should be further considered and can reduce the room-temperature thermal conductivity by ~13% at a doping level of 7.0×10$^{18}$ cm$^{-3}$.[16] This new phonon-scattering mechanism is further considered for heavily doped GaN nanowires and the scattering rate is given as[12]

$$\tau_E^{-1}(\omega)=\frac{D_a^2 m^{*3} v_g}{4\pi\hbar^4\rho}\frac{k_B T}{\frac{1}{2}m^* {v_g}^2}\times\left\{\frac{\hbar\omega}{k_B T}-\ln\frac{1+\exp\left[\left(\frac{1}{2}m^* {v_g}^2-E_F\right)/k_B T+\hbar^2\omega^2/8m^* {v_g}^2 k_B T+\hbar\omega/2k_B T\right]}{1+\exp\left[\left(\frac{1}{2}m^* {v_g}^2-E_F\right)/k_B T+\hbar^2\omega^2/8m^* {v_g}^2 k_B T-\hbar\omega/2k_B T\right]}\right\}, \quad (1)$$

where $T$, $\hbar$, $k_B$, $D_a$, $m^*$, $\rho$, $v_g$, $E_F$ represent absolute temperature, Planck constant divided by 2π, Boltzmann constant, acoustic deformation potential, density of states (DOS) effective mass, density, averaged phonon group velocity,[17] and Fermi level, respectively.

Special attention should be paid to the phonon transport across an interface. In particular, it is known that the GaN-substrate thermal boundary resistance plays an important role in restricting heat spreading.[18,19] Based on the diffuse mismatch model, phonons are diffusively transmitted or reflected by an interface. The frequency-dependent phonon transmissivity from material 1 to 2 is given as[20]

$$\tau_{12}(\omega)=\sum_p v_{2,g,p}(\omega)D_{2,p}(\omega)/\left[\sum_p v_{1,g,p}(\omega)D_{1,p}(\omega)+\sum_p v_{2,g,p}(\omega)D_{2,p}(\omega)\right], \quad (2)$$

in which the subscript 1 or 2 indicates the material, subscript $p$ indicates the phonon branch, $\omega$ is the phonon angular frequency, $D_p(\omega)$ is the phonon DOS for branch $p$, and $v_{g,p}(\omega)$ is phonon group velocity for branch $p$. Detailed treatment of interfacial scattering in phonon MC simulations can be found elsewhere.[21,22]

Using $\tau_{12}(\omega)$, the interfacial thermal resistance $R_K$ can then be computed as[20]

$$1/R_K=\sum_p\left[\int_0^{\omega_{1,\max,p}} v_{1,g,p}(\omega)c_{1,p}(\omega)\tau_{12}(\omega)d\omega\right]/4, \quad (3)$$

with $\omega_{1,\max,p}$ as the maximum $\omega$ value for branch $p$ in material 1. This $R_K$ value is used in the Fourier's law analysis to be consistent with phonon MC simulations. To improve the accuracy of the Fourier's law analysis, the bulk phonon MFPs within each nanostructure or microstructure is modified using the structure size, such as the film thickness and nanowire diameter.[23]

Figure 1 presents the simulated FinFETs using an array of 31 parallel Si-doped GaN nanowires. The large nanowire array is used to obtain a large total output current. To reduce the computational load, one nanowire in the middle is chosen for the study and the phonon MC domain is indicated by dashed lines. The distance from the hot spot to the boundary of this domain is 6–10 μm, which is longer than majority phonon MFPs in GaN and SiC[11] to validate the Fourier's law analysis outside the phonon MC domain. On planes A and B, specular phonon reflection is enforced due to structure symmetry[21] within such a large nanowire array. Although this boundary condition is less accurate for the nanowires on the edge of the nanowire array, limited influence is anticipated for a nanowire in the middle of the array and the computational load can be largely minimized with the proposed computational domain. In the hybrid simulations, electron MC simulations predict local phonon emission by hot electrons along the nanowire, which is input into the phonon MC simulations and the Fourier's law analysis as the heat generation. The phonon MC simulations and Fourier's law analysis are then carried out to refine the temperature predictions across the whole chip. The phonon MC simulations particularly update the local temperature along the nanowire, which affects the local electron scattering rates in electron MC simulations. The three simulations are carried out in an iterative way until the steady-state temperature distribution is obtained for the GaN FinFET. More details of this simulation technique can be found in our previous studies.[6,7]

Figures 2a and 2b show the predicted acoustic phonon temperature distribution along the nanowire, both in a top view and a side view. The source, drain, and gate are at x=6–6.5 μm, 8.5–9 μm, and 7–8 μm, respectively. Here the source and gate are both grounded, while 10 V voltage is applied to the drain. The peak electric field and thus peak temperature occur at the drain-side gate edge, similar to that observed in conventional planar GaN/AlGaN device.[7] The temperature distribution along the middle nanowire, computed by the Fourier's law analysis, is also plotted in comparison with that from phonon MC simulations (Fig. 3). Both temperature distributions converge at the boundary of the phonon MC domain. This comparison can be used to justify the size of the selected phonon MC domain.

In practice, the phonon MC domain cannot include all nanowires due to the huge computational load. The heat generation and boundary condition are anticipated to be different for nanowires from the middle to the edge of the nanowire array. One concern is whether different nanowires may have different temperature profiles and thus heat generation by hot electrons. To check the temperature variation across different nanowires, the temperature profiles along the 1$^{st}$ (edge), 8$^{th}$, and 16$^{th}$ (middle) nanowires in the array are computed with the Fourier's law (Fig. 4). For the nanowire on the edge of the array, more heat leaks into the GaN film so that the maximum temperature rise is ~25 K lower than that for the middle nanowire. The Fourier's law predictions are lower than the accurate temperature rise given by phonon MC simulations due to strong ballistic phonon transport in nanowire devices. Such temperature underestimation is anticipated to be similar for different nanowires so that the actual acoustic phonon temperature rise is within ~25 K divergence among all nanowires, i.e., within 8% divergence of the absolute temperature. The corresponding electron scattering rates by phonons are then very close along different nanowires, leading to almost identical heat generation. Therefore, it is reasonable to assume approximately equal heat generation for all nanowires, as extracted from electron MC simulations for the middle nanowire.

As another important parameter in device thermal studies, the device thermal resistance is also computed as the maximum temperature rise divided by the total power dissipation of 31 identical nanowires. Using the Fourier's law, previous calculations for the device thermal resistance[19] only consider acoustic phonons and completely neglect the thermal non-equilibrium between hot electrons, optical phonons, and acoustic phonons. In addition, heat dissipation is often computed as the classical Joule heat, which is the dot product between the current density and electric field. This is again inaccurate because hot electrons usually travel a few of their MFPs before passing their energy to phonons. Considering these issues, the device thermal resistance is redefined as the maximum acoustic phonon temperature rise $\Delta T_{ac}$ divided by the total energy $E_{\mathrm{Emit}}$ of emitted phonons, as counted in electron MC simulations. Different drain voltages are used in the computations and the computed device thermal resistance is plotted against the applied drain voltage in Fig. 5. Compared with calculations using the Fourier's law, $\Delta T_{ac}$ from phonon MC simulations tends to predict a higher device thermal resistance. The large difference is partially attributed to the ballistic thermal resistance between the nanowire and GaN film.[24]

Phonons with MFPs much longer than the ~100 nm width of the nanowire can travel ballistically into the GaN film. This leads to largely reduced heat transfer compared to the Fourier's law. In bulk GaN, ~60% of the lattice thermal conductivity at 415 K is contributed by phonons with MFPs longer than 100 nm but less than the 2 μm thickness of the GaN film.[11] Therefore, a large ballistic thermal resistance is anticipated at nanowire-film junctions. Despite many advantages of nanostructured devices, such ballistic thermal resistance should receive more attention for its resulting increased temperature rise within the device.

In summary, electrothermal simulations have been carried out to understand the heat generation and transport within nanowire-based FinFETs. The coupled electron and phonon MC simulations allow accurate predictions for both the temperature rise and device characteristics. For nanostructured devices, the ballistic thermal resistance at the nanostructure-substrate interface can be critical, as observed in nanowire-based FinFETs here. In this situation, thermal management by improved cooling on the substrate side is limited for reducing the hot-spot temperature. Coating the top of the device with a high-thermal-conductivity heat spreader can be more effective.

The authors thank Prof. David Broido for the input on the phonon relaxation times within bulk GaN; also thank Kevin Bagnall and Prof. Evelyn Wang for help with ATLAS simulation. This material is based on research sponsored by Defense Advanced Research Agency (DARPA) under agreement number FA8650-15-1-7523. Some phonon studies are also supported by National Science Foundation under grant number CBET-1651840. The U.S. Government is authorized to reproduce and distribute reprints for Governmental purposes notwithstanding any copyright notation thereon. The views and conclusions contained herein are those of the authors and should not be interpreted as necessarily representing the official policies or endorsements, either expressed or implied, of Air Force Research Laboratory (AFRL) and the DARPA or the U.S. Government. An allocation of computer time from the UA Research Computing High Performance Computing (HPC) and High Throughput Computing (HTC) at the University of Arizona is gratefully acknowledged.

**Figure Captions**

Fig. 1 Schematics of the simulated 3D GaN-on-SiC device: (a) 3D structure, and (b) cross-sectional view perpendicular to the nanowires.

Fig. 2 Temperature profiles from the coupled electrothermal simulation of GaN-on-SiC FinFETs: (a) side view at mirror symmetry plane A in Fig. 1; and (b) top view of the nanowire array. All temperatures are in Kelvin.

Fig. 3 (a) Difference between the temperature profiles predicted by the phonon MC simulation and Fourier's law analysis for the middle nanowire. (b) Temperature along the middle nanowire based on the phonon MC simulation (solid line) and Fourier's law analysis (dashed line).

Fig. 4 Temperature profiles along 1$^{st}$, 8$^{th}$, and 16$^{th}$ nanowires, as computed with the Fourier's law.

Fig. 5 Device thermal resistance for the whole nanowire array as a function of the applied drain voltage, assuming grounded source and gate.

**Tables**

TABLE I. Parameters used for bulk GaN in electron MC simulations.

| Parameter (Unit) | Symbol | Value |
|---|---|---|
| Electron density ($cm^{-3}$) | $n$ | $4.0\times10^{18}$ |
| Electron effective mass ($m_0$) | $m_i^*$ | 0.21, 0.25, 0.40 |
| Valley minimum energy (eV) | $E_{c,i}$ | 0, 1.95, 2.1 |
| Nonparabolicity ($eV^{-1}$) | $\alpha_i$ | 0.19, 0.1, 0 |
| Dielectric constant ($\varepsilon_0$) | $\varepsilon_s$, $\varepsilon_\infty$ | 8.9, 5.35 |
| Mass density ($g/cm^3$) | $\rho$ | 6.095 |
| Acoustic deformation potential (eV) | $D_a$ | 8.3 |
| Intervalley deformation potential (eV/cm) | $D_{ij}$ | $1.0\times10^9$ |

TABLE II. Fitted phonon dispersion and scattering parameters for materials in a FinFET.

| Parameter (Unit) | GaN | 6H-SiC |
|---|---|---|
| $k_0$ ($10^9$ $m^{-1}$) | 10.94 | 8.94 |
| $\omega_{\max}$ ($10^{13}$ rad/s) | 3.50 | 7.12 |
| $a_D$ (Å) | 2.87 | 3.51 |
| A ($10^{-45}$ $s^3$) | 5.26 | 1.00 |
| $B_1$ ($10^{-19}$ s/K) | 1.10 | 0.596 |
| $B_2$ (K) | 200.0 | 235.0 |

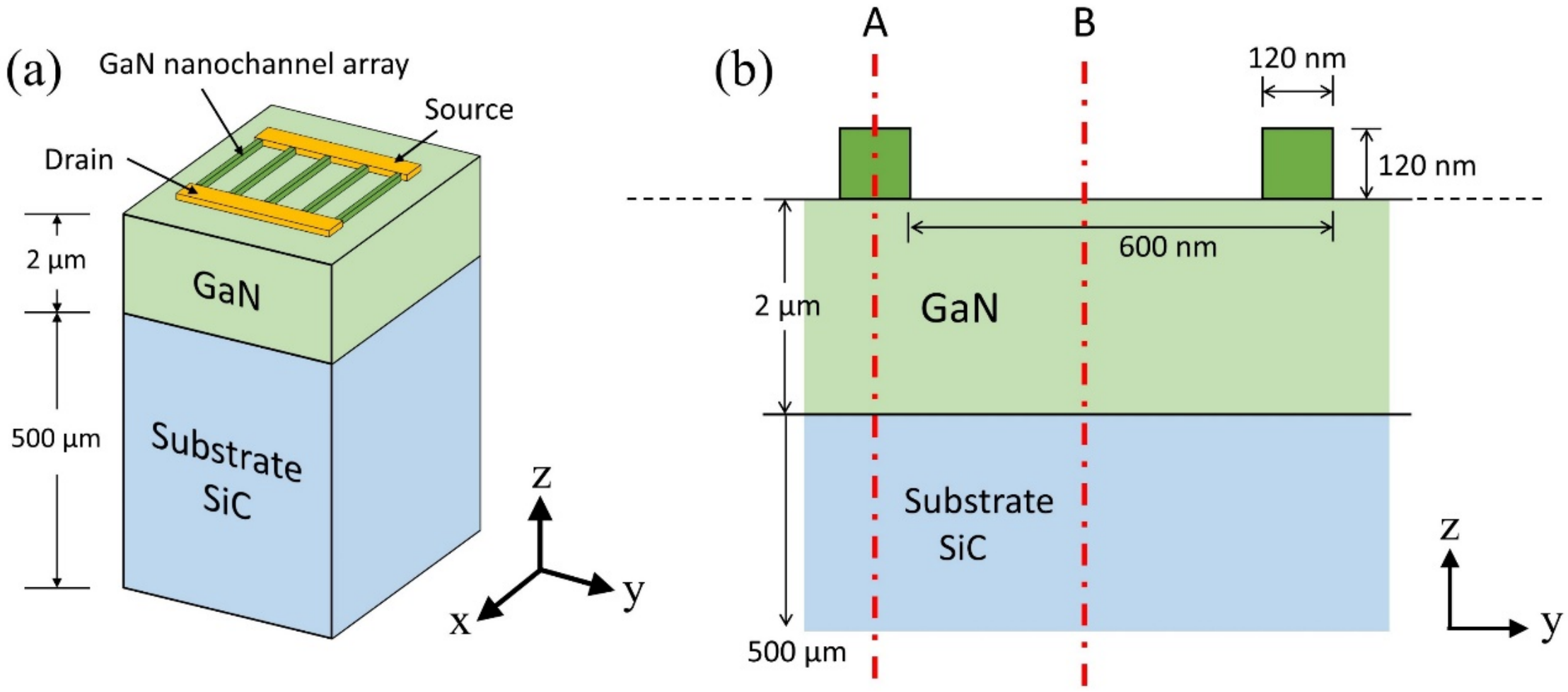

Fig. 1. Hao et al.

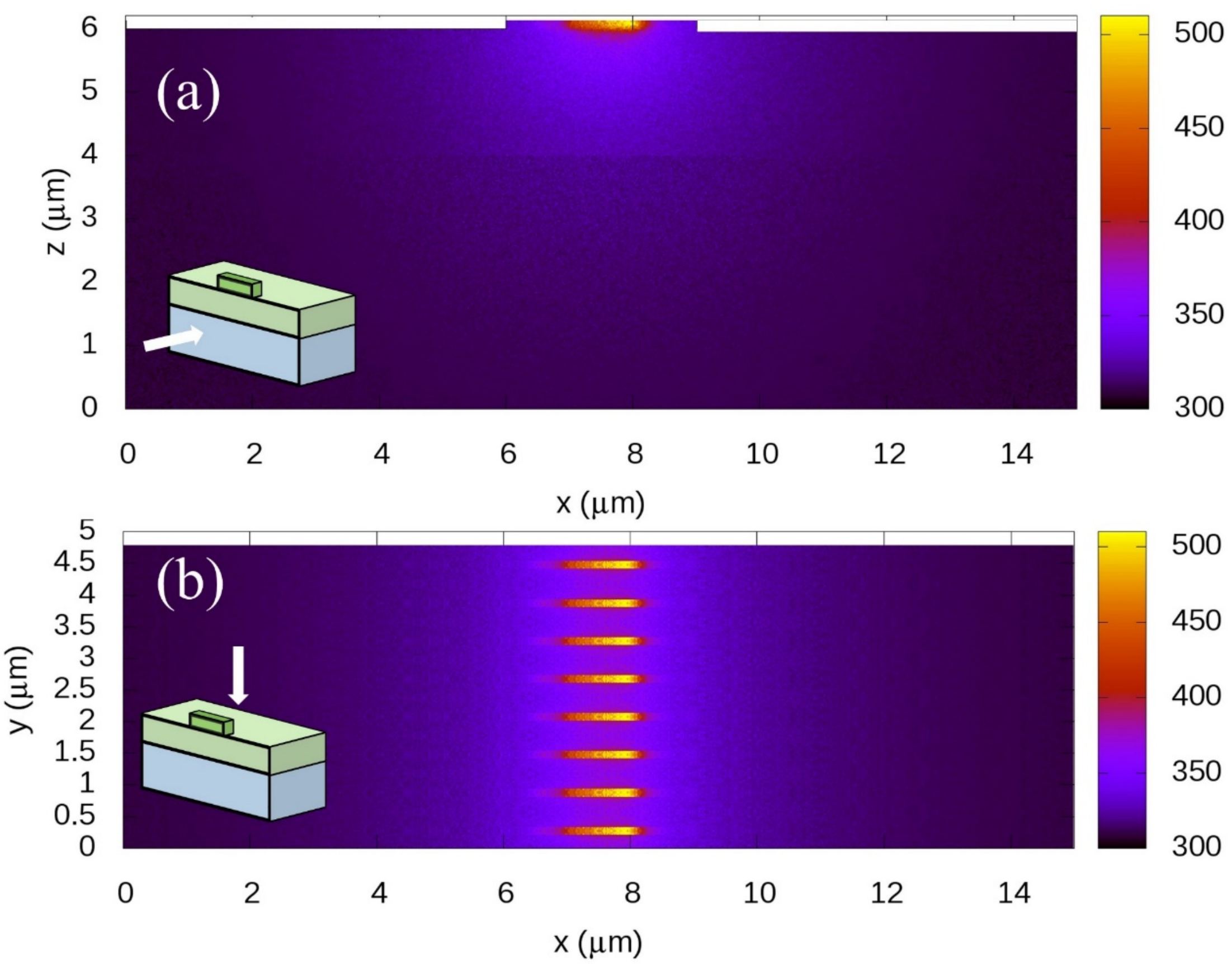

Fig. 2. Hao et al.

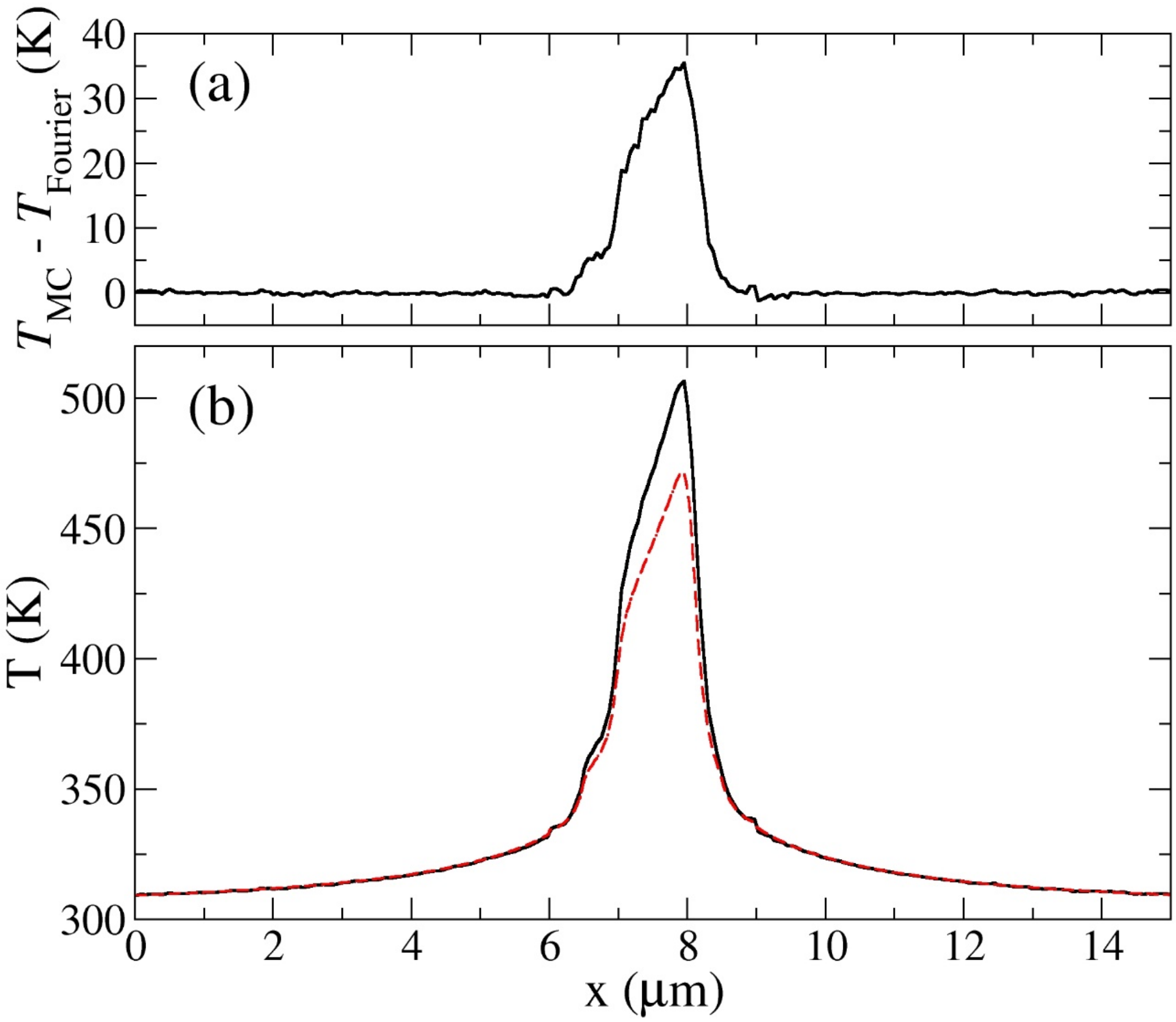

Fig. 3. Hao et al.

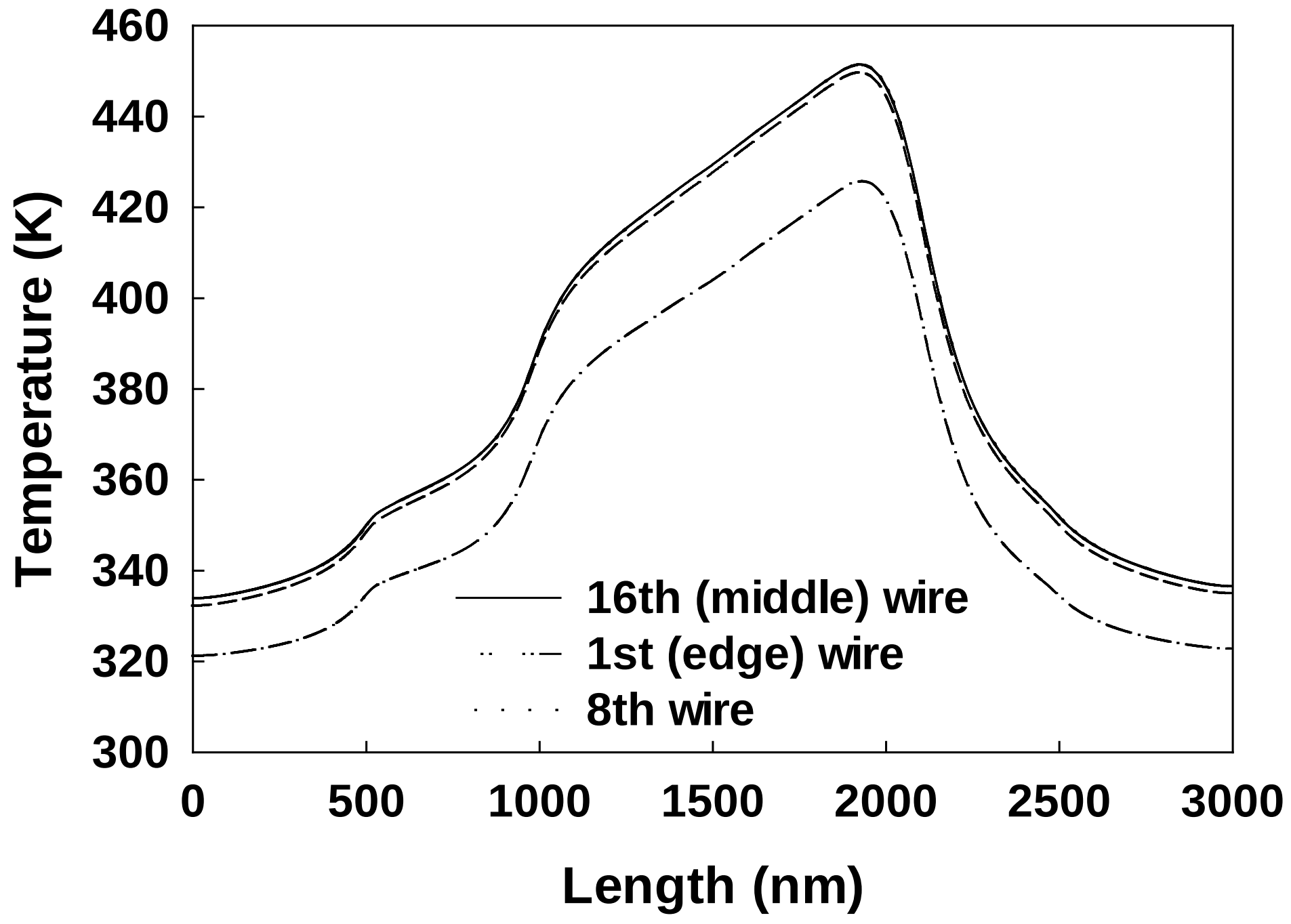

Fig. 4. Hao et al.

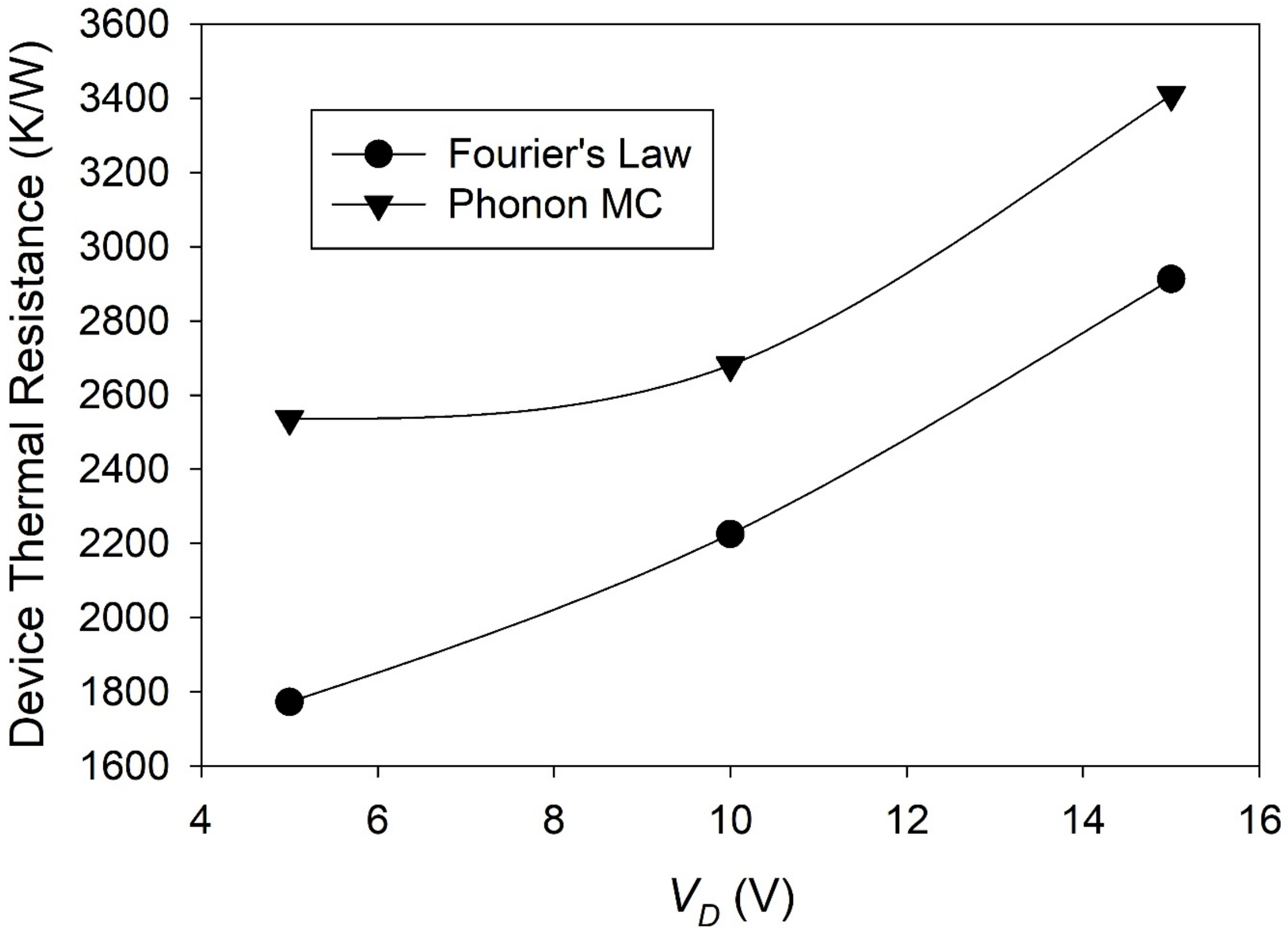

Fig. 5. Hao et al.